# Impact cratering on porous targets in the strength regime

Akiko M. Nakamura

Graduate School of Science, Kobe University, 1-1 Rokkodai-cho, Nada-ku, Kobe 657-8501, Japan

Corresponding author:
Akiko M. Nakamura
TEL : +81-78-803-5740
FAX: +81-78-803-5791
e-mail: amnakamu@kobe-u.ac.jp

Table 1
Figures 6

ABSTRACT

Cratering on small bodies is crucial for the collision cascade and also contributes to the ejection of dust particles into interplanetary space. A crater cavity forms against the mechanical strength of the surface, gravitational acceleration, or both. The formation of moderately sized craters that are sufficiently larger than the thickness of the regolith on small bodies, in which mechanical strength plays the dominant role rather than gravitational acceleration, is in the strength regime. The formation of microcraters on blocks on the surface is also within the strength regime. On the other hand, the formation of a crater of a size comparable to the thickness of the regolith is affected by both gravitational acceleration and cohesion between regolith particles.

In this short review, we compile data from the literature pertaining to impact cratering experiments on porous targets, and summarize the ratio of spall diameter to pit diameter, the depth, diameter, and volume of the crater cavity, and the ratio of depth to diameter. Among targets with various porosities studied in the laboratory to date, based on conventional scaling laws (Holsapple and Schmidt, *J. Geophys. Res.*, 87, 1849-1870, 1982) the cratering efficiency obtained for porous sedimentary rocks (Suzuki et al., *J. Geophys. Res.* 117, E08012, 2012) is intermediate. A comparison with microcraters formed on a glass target with impact velocities up to 14 km $s^{-1}$ indicates a different dependence of cratering efficiency and depth-to-diameter ratio on impact velocity.

## 1. INTRODUCTION

Mass ejection by cratering has been found to be crucial in collision cascades which are one of the most important processes in the main asteroid belt, the Edgeworth-Kuiper Belt, debris disks, and planet formation (Kobayashi and Tanaka, 2010). Cratering on small bodies contributes to the ejection of dust particles into interplanetary space (Yamamoto and Mukai, 1998; Tomeoka et al., 2003). During high-velocity impact on the surface of a planetary body, a crater cavity is excavated against the mechanical strength of the surface, the gravitational acceleration, or both (Holsapple, 1993). When the surface has a porous structure, cavity formation also proceeds by compaction of that structure (Housen and Holsapple, 2003). In other words, the response of a planetary surface to impact depends on its physical properties, such as mechanical strength, porosity, and internal structure (Nakamura et al., 2009), and on its gravitational acceleration.

The size of a relatively large crater is limited mainly by gravitational acceleration, rather than by mechanical strength. This situation is called the gravity regime. Even when a small crater is created, if the mechanical strength is negligible, such as when a millimeter-size projectile impacts a sand target in the laboratory, the size of the crater cavity is dependent mainly on gravitational acceleration. On the other hand, when the effect of gravity is negligible compared with the mechanical strength, the impact process and its consequences, such as the dimensions and shape of the cavity, are in the strength regime and depend on the material strength, porosity, and structural characteristics such as the density structure or spatial inhomogeneity. In the laboratory, the speed of cratering ejecta has been shown to be related to the mechanical strength of the surface, i.e., the stronger the target, the higher the ejecta speed (Housen, 1992; Michikami et al., 2007; Housen and Holsapple, 2011). Cratering efficiency is affected by the size ratio between the projectile and the target grain when the projectile size is so small that the disruption of the single target grain by the projectile is not negligible (Guettler et al., 2012).

The appearance of craters on small bodies differs markedly among different types of bodies. Large craters, with diameters comparable to the diameter of the asteroid, coexist on the asteroid 253 Mathilde (Veverka et al., 1999). The reason why such large craters can coexist, and why the body was not disrupted or dispersed, is probably that the porosity effectively attenuated the shock pressure via compaction (Housen et al., 1999; Housen and Holsapple, 2003). Ejecta blocks of tens of meters to ~100 m in size have been found in the vicinity of craters on Martian moons (Lee et al., 1986) and asteroid 243 Ida (Lee et al., 1996), and are even widespread over the surface of asteroid 433 Eros (Thomas et al., 2001). However, no ejecta blocks have been identified on Mathilde (Veverka et al., 1999), this is attributed to the high porosity of the body (Housen and Holsapple, 2012). Ejecta blocks that are meters to tens of meters in size are more prominent on the rubble-pile asteroid 25143 Itokawa (Michikami et al., 2008), where typical bowl-shaped craters have not been identified but very shallow circular depressions have been observed (Hirata et al., 2009).

As shown in Section 2, craters that are tens to hundreds of meters in size on a ~1-km diameter body such as asteroid 162173 Ryugu, the mission target of Hayabusa2, can be affected markedly by the mechanical strength, porosity, and structure of the surface. Microcraters on blocks and even on the regolith particles on the surface of small bodies also form in the strength regime. Most small solar system bodies are porous, which means, that the bulk density of the bodies is smaller than those of the component materials of the bodies such as chondrites (Consolmagno et al., 2008). The S-type asteroid Itokawa has been shown to be a rubble-pile, i.e., a re-accumulated body, and has a macroporosity of 40% according to the results from Hayabusa (Fujiwara et al., 2006). The C-type asteroid Mathilde has also been explored by the space mission NEAR Shoemaker. The bulk density of this body, 1300 kg $m^{-3}$, shows that its macroporosity is about 50% (Veverka et al., 1999). Moreover, chondrites are also porous, i.e., ordinary chondrites have a porosity of several to tens of percent and carbonaceous chondrites are more porous (Consolmagno et al., 2008).

Here we review laboratory high-velocity impact experiments performed to study the dimensions and shapes of craters on brittle targets, especially on porous targets. In Section 2, the size of craters expected to be produced in the strength regime is roughly defined. In Section 3, laboratory experiments on the dimensions and shapes of craters on brittle targets, with a focus on porous targets, are summarized. Recent laboratory experiments have been conducted using millimeter- to centimeter-size projectiles, resulting in craters ranging in diameter from centimeters to tens of centimeters. However, microcraters have been found on lunar samples, as well as on samples returned from asteroid Itokawa (Nakamura et al., 2012; Matsumoto et al., 2016; Harries et al., 2016); therefore, craters with diameters smaller than 1 mm are also of interest. Hence, we also refer to the results of previous laboratory impact experiments investigating microcraters and make a comparison between microcraters and centimeter-scale craters formed in the laboratory.

## 2. SIZE OF CRATERS ON SMALL BODIES IN THE STRENGTH-DOMINATED REGIME

The gravity-strength transition diameter of the crater regime can be estimated roughly by balancing the mechanical strength of the surface $Y$ and the lithostatic pressure, $\rho g l$, where $\rho$, $g$, and $l$ are the density of the surface, the gravitational acceleration, and the representative scale in terms of crater dimensions, respectively. If we take the crater diameter as the representative scale, the transition diameter $D_{tr}$ satisfies the relationship

$$Y = k_1 \rho g D_{tr}, \quad (1)$$

where $k_1$ is a factor. Substituting $g = G\rho\frac{4\pi}{3}R$, where $R$ is the radius of a spherical body, yields

$$Y = k_1 \rho \left(G\rho\frac{4\pi}{3}R\right) D_{tr}. \qquad (2)$$

Accordingly, the relationship of the transition diameter versus the radius of the body is given as

$$D_{tr} = \frac{3Y}{4k_1 \pi G \rho^2 R} \ . \quad (3)$$

Note that a thorough analysis of the cratering process based on a point source assumption shows that the transition between strength and gravity regimes depends on the impact velocity and occurs at $Y \approx \rho g a$, where $a$ is the impactor radius (Holsapple, 1993) and derives a functional form for the transition involving impact velocity (e.g., Housen and Holsapple, 2011).

Figure 1 shows the crater diameter of the gravity-strength transition according to Eq. 3. For example, craters smaller than ~18 km in diameter on an 18-km-diameter body having a mechanical strength of 0.1 MPa and density of 1500 kg $m^{-3}$ can form in the strength regime, if $k_1$ is unity or less. However, the surface of a body is generally covered by regolith. For example, the thicknesses of the regolith on Eros and Phobos have been estimated to be tens of meters (Thomas et al., 2001) and 5-100 m (Basilevsky, et al., 2014), respectively. Given that the nominal depth-to-diameter ratio of a bowl-shaped crater is 0.2 (Pike, 1974), craters with diameter $D$ in the following range can be considered to have formed in the strength regime.

$$5h \ll D < D_{tr} \text{ , (4)}$$

where $h$ is the thickness of regolith layer.

Regolith particles are produced by impact (Housen and Wilkening, 1982) and by thermal fatigue (Delbo et al., 2014). Here, we consider only production by impact. To obtain a first-order estimate of regolith thickness, we assume that impact cratering occurs on a consolidated surface and fragments the surface materials, and that a portion of the disaggregated materials re-accumulates and becomes distributed uniformly over the entire surface of the spherical body. The size distribution of the ejecta particles and the erosion of pre-existing regolith materials by impact are not taken into consideration. Then the average regolith thickness $h(D)$ owing to a crater of diameter $D$ is given by

$$h(D) = \frac{M(D)}{4\pi R^2 \rho_{regolith}}, \text{ (5)}$$

where $M(D)$ is the mass of re-accumulated ejecta and $\rho_{regolith}$ is the bulk density of the regolith layer. The bulk density of the regolith layer is generally lower than the bulk density of the body, i.e., $\rho_{regolith} \leq \rho$.

We assume a relationship between the volume of fragmented materials $V(D)$ and diameter $D$ as follows:

$$V(D) = k_2 D^3 \text{ , (6)}$$

where $k_2$ is a constant. We assume $k_2$ = 0.07, which provides a value that corresponds approximately to the largest of that obtained experimentally for small laboratory craters shown in Section 3.4. In a previous study, the volume of materials ejected from craters was assumed to be ~$0.06D^3$ based on the shape of fresh-looking craters on asteroid Ida (Lee et al., 1996).

The re-accumulated fraction depends on the velocity distribution of the ejecta and the gravitational acceleration of the body. When the crater forms in the strength regime, the total mass of ejecta having speed larger than $v$, $M(>v, D)$ normalized by the mass of some representative volume $\left(\frac{D}{2}\right)^3$ (not exactly the cavity volume) is given by

$$\frac{M(>v,D)}{\rho(\frac{D}{2})^3} = f_s\left(\frac{v}{\sqrt{Y/\rho}}\right) \quad (7)$$

(Housen, 1992; Housen and Holsapple, 2011). The empirical relationship between non-dimensional ejecta speed and mass fraction and then the function $f_s\left(\frac{v}{\sqrt{Y/\rho}}\right)$ has been determined experimentally for basalt (Housen, 1992), sintered glass bead targets (Michikami et al., 2007; Michikami et al., 2008), and a weak cemented basalt and perlite–sand mixture (Housen and Holsapple, 2011). In these studies, tensile strength was used for $Y$. The function $f_s$ decreases with $v$ and has a power law form for the ejecta except for those from the impact point (the highest ejection speed component) and those from the crater rim (the lowest ejection speed component):

$$f_s\left(\frac{v}{\sqrt{Y/\rho}}\right) \approx C_s\left(v\sqrt{\frac{\rho}{Y}}\right)^{-\beta_s}, \quad (8)$$

where $C_s$ and $\beta_s$ are positive constants; within the theoretical framework of the point source approximation of crater formation we require that $1 < \beta_s < 2$ (e.g., Housen and Holsapple, 2011).

Thus, the regolith thickness $h(D)$ owing to a crater of diameter $D$ is given by

$$h(D) = \frac{\rho k_2 D^3 - M(>v_{escape})}{4\pi R^2 \rho_{regolith}}, \quad (9)$$

where $v_{escape}$ is the escape velocity of the body and equals to $R\sqrt{\frac{8}{3}\pi\rho G}$ for a spherical homogeneous body. Assuming $\rho_{regolith} = \rho$, then

$$h(D) = \frac{D^3}{32\pi R^2}\left\{8k_2 - f_s\left(v_{escape}\sqrt{\frac{\rho}{Y}}\right)\right\} = \frac{D^3}{32\pi R^2}\left\{8k_2 - f_s\left(\rho R\sqrt{\frac{8\pi G}{3Y}}\right)\right\}$$

$$\approx \frac{D^3}{32\pi R^2}\left\{8k_2 - C_s\left(\rho R\sqrt{\frac{8\pi G}{3Y}}\right)^{-\beta_s}\right\}. \quad (10)$$

Because of the size distribution of the possible craters (or the possible impactors), the ejecta from the largest craters are expected to be dominant in the total volume of the regolith. In particular, the volume of the ejecta from the largest crater comprises more than half (Thomas et al., 2001), or a fraction (Michikami et al., 2008) of the total volume of the ejecta from the largest craters. Therefore, we examine the contribution of the largest possible crater. An empirical relationship of the largest crater diameter $D_{max}$ on a small body is given below:

$$D_{max} = k_3 R, \quad (11)$$

where $k_3$ depends on the internal structure of the body and is approximately unity (Burchell and Leliwa-Kopystynski, 2010). Figure 1 shows $h(D_{max})$ for two types of targets. We assume $k_3$=1. We also show the case in which all of the fragmented materials re-accumulate on the surface to contribute to the thickness of the regolith. The expected thickness of the regolith layer is very thin for a body with diameter of less than a few kilometers. The average total thickness of the regolith would be larger than $h(D_{max})$ by some factors. Craters with diameters below the gravity-strength transition and the maximum crater diameter, $D_{max}$, and well above the thickness of regolith owing to the largest crater, $h(D_{max})$, are formed in the strength regime. For example, tens to hundreds of meters-sized craters on a ~1-km-diameter body are formed in the strength regime.

However, craters on the surface of a block or on a regolith particle are strength regime craters even when they have diameter much smaller than the regolith thickness.

## 3. HIGH-VELOCITY IMPACT CRATERING EXPERIMENTS OF POROUS TARGETS IN THE STRENGTH REGIME

Laboratory high-velocity impact experiments, using millimeter-to-centimeter size projectiles to study cratering in the strength regime, were performed for various targets with porosities ranging from 0% to more than 90%. The morphology and dimensions, i.e., diameter and depth, were studied for weak cemented basalt targets (Housen, 1992), sand–perlite–fly ash mixture targets (Housen and Holsapple, 2003), sintered glass bead targets (Love et al., 1993; Michikami et al., 2007; Hiraoka et al., 2008; Okamoto et al, 2015; Okamoto and Nakamura 2017), cement mortar targets (Michikami et al., 2017), gypsum targets (Yasui et al., 2012; Okamoto and Nakamura, 2017), and natural porous rocks (Baldwin, et al., 2007; Kenkmann, et al., 2011; Suzuki et al., 2012; Poelchau, M. H., 2014; Flynn, et al., 2015; Okamoto and Nakamura, 2017). Table 1 summarizes the target density, porosity, strength, impact velocity, and projectile material and diameter of these experiments. Compaction of porous targets was reported for sand–perlite–fly–ash mixture (Housen and Holsapple, 2003), gypsum (Yasui et al., 2012), sandstone (Buhl et al., 2013), and tuff targets (Winkler et al., 2016).

### 3.1 Spall and pit diameters

Craters formed on solid targets with relatively low porosity in the laboratory have a bowl-shaped cavity (pit) with a surrounding thin depression (spall). The shape of the cavity for relatively porous targets is cylindrical, U-shaped or bulb-shaped (Michikami et al, 2007; Flynn et al., 2015; Winkler et al., 2016; Okamoto and Nakamura, 2017). The spall zone is not identified on highly porous targets, such as a sintered glass bead target with 80% porosity, but occurs for those with 60% porosity (Michikami et al., 2007). The spallation occurs due to tensile failure when the compressive stress is reflected at the surface and becomes tensile stress (Melosh et al., 1989). It was reported that all microcraters larger than 50 μm on lunar rocks had spall (Hoerz et al., 1975). A power-law relationship $D_{spall} = 2.37{D_{pit}}^{1.07}$ was obtained for data from lunar microcraters with spall diameters ranging from 8.6 to 390 μm, where $D_{spall}$ and $D_{pit}$ are the spall and pit diameters in microns, respectively (Morrison et al., 1973). The relationship shows that the spall-to-pit diameter ratio for lunar microcraters is ~3. Such microcraters with spall were reproduced on glass targets by laboratory impact experiments (Mandeville and Vedder, 1971).

Figure 2 summarizes the data regarding spall-to-pit diameter ratio versus porosity of target. There is no strong tendency with porosity. This is because the spall diameter and pit diameter correspond to the stress level of the tensile strength and shear or compressive strength of the target material, respectively, and because the ratio of the tensile strength to the shear strength (or compressive strength) does not vary much among the different target materials, as shown in Table 1. The ratio of the spall diameter to the pit diameter of porous targets ranges between 1.5 and 3. The latter value is the average of the lunar microcraters mentioned above. The data on laboratory microcraters on glass targets (Mandeville and Vedder, 1971) fall in a range similar to those of the porous targets in Fig. 2.

## 3.2 Depth of crater

Crater depth on porous targets has been shown to be dependent on the ratio of impactor density $\delta$ to target density $\rho$ (Love et al., 1993; Michikami et al., 2007; Okamoto and Nakamura, 2017). This is because the penetration depth of the impactor increases when the density ratio of the impactor to the target increases. Quantitatively, the equivalent depth of burial of an explosive $d_b$ for an impact has been shown to depend on the density ratio, as given by the following equation (Melosh, 1989):

$$\frac{d_b}{2r_p} = \left(\frac{\delta}{\rho}\right)^{0.5} , \qquad (12)$$

where $r_p$ is the impactor radius. On the other hand, the cavity depth of porous targets including highly porous aerogels increases more rapidly with the density ratio (Love et al., 1993; Michikami et al., 2007). A semi-empirical expression for the penetration depth $d_p$ of an impactor for highly porous targets with a linear dependence on the density ratio has been proposed as follows (Okamoto et al., 2013):

$$\frac{d_p}{2r_p} = \frac{2}{3C_d}\left(\frac{\delta}{\rho}\right) \ln(1 + C_d \frac{\rho U^2}{2Y_c}) \times \min(1, \{10^{1.5\pm0.7} \left(\frac{\rho U^2}{Y_{tp}}\right)^{-1.5\pm0.5} \}^{\frac{1}{3}}) , \qquad (13)$$

where $C_d$, $Y_c$, $Y_{tp}$, and $U$ are the effective drag coefficient for the impactor when it penetrates a porous target, compressive strength of the target, tensile strength of the impactor, and impact velocity, respectively. An intermediate dependence on the density ratio, i.e., a power law dependence with an index between 0.5 and 1 was obtained for crater depth, $d$ of porous targets of foamed polystyrene, sintered glass bead, gypsum, and pumice by curve fitting (Okamoto and Nakamura, 2017):

$$\frac{d}{2r_p} = 10^{0.46\pm0.03} \left(\frac{\delta}{\rho}\right)^{0.72\pm0.03} , \qquad (14)$$

although the data points scattered around the relationship given by Eq. 14. The compaction of a target and deeper penetration of an impactor are possible sources of the larger power index of Eq. 14 versus that of Eq. 12. Figure 3 shows the literature data for porous targets. The empirical relationship given by Eq. 14, and some of the data (gypsum and pumice data) used in the derivation of the relationship, are also shown. Considering the scatter of gypsum and pumice data points across the empirical relationship, it can be said that the overall tendency especially of craters formed by high-velocity impactors ($U > 4$ km s$^{-1}$), is close to Eq. 14. However, the figure illustrates that cavity depth is also dependent on impact velocity, i.e., the higher the impact velocity, the greater the depth normalized by the diameter of the impactor. It also indicates that more porous targets such as tuff, gypsum, and pumice, result in deeper cavities than in sandstones.

### 3.3 Cratering efficiency

The diameter of a crater formed on a porous targets tends to be smaller than that of one on a non-porous target, due to stronger attenuation of stress waves in porous targets. However, on the other hand, because more-porous targets are weaker, the crater may be larger on fragile porous targets. Such complexity was observed for a sintered glass bead target of 37% porosity and a 39% target (Love et al., 1993). The total ejecta mass and volume were smaller for the 37% target than for the 39% target. The former had strength four times the strength of the latter. Baldwin et al. (2007), Suzuki et al. (2012), and Poelchau et al. (2014) compiled results of porous rock targets using conventional cratering scaling laws (Holsapple and Schmidt, 1982) and showed that the cratering efficiency, defined by the ratio of excavated mass to projectile mass, for the porous targets was less than that of competent rocks for the same impact conditions defined by a non-dimensional parameter involving the mechanical strength of the target, such as

$$\pi_3 = \frac{Y}{\delta U^2}. \qquad (15)$$

The lower cratering efficiency indicates more rapid attenuation of the stress wave in porous sedimentary rocks. The relationship between the normalized crater diameter $\pi_D = \left(\frac{\rho}{m}\right)^{\frac{1}{3}} D$, where $m$ denotes projectile mass, and the non-dimensional impact condition obtained by Suzuki et al. (2012) is

$$\pi_D = H_D \pi_3^{-\beta_D} \pi_4^{\gamma_D}, \qquad (16)$$

where

$$\pi_4 = \frac{\rho}{\delta}, \qquad (17)$$

and

$$H_D = 1.43 \pm 0.25,\ \beta_D = 0.22 \pm 0.02,\ \text{and}\ \gamma_D = 0.11 \pm 0.07. \ (18)$$

The tensile strength is used for $Y$ in $\pi_3$. Figure 4a shows the literature data for porous targets with known tensile strengths along with Eqs. 16 and $\gamma_D = 0.11$. Here we use the spall diameter for $D$, except for pumice data. The general tendency is that the cratering efficiency of more porous targets, such as gypsum (50% porosity) and tuff (43%), is lower than that of sandstones (~20%), and much lower than that of non-porous basalt. However, the data for weak cemented basalt (23%) fall not in the area of sandstones but on the trend of gypsum. On the other hand, the slope of the microcraters on glass is similar to the empirical relationship for sedimentary rock (PS and CS) and shallower than that of the craters on basalt targets. The difference in slope between microcraters on a glass target and centimeter craters on basalt targets is probably due to differences in impact velocity and target strength.

An alternative expression for crater diameter using a non-dimensional parameter $\frac{Y}{\rho U^2}$ instead of $\pi_3$ is given as

$$\pi_D = {H_D}^* \left(\frac{Y}{\rho U^2}\right)^{-\beta_D^*} \pi_4^{\gamma_D^*}. \ (19)$$

The proportional constant $H_D^*$ and the power index $\beta_D^*$ are obtained for porous targets consisting of weak cemented mortar, fly ash–sand mixture and perlite–sand mixture targets with the value of $\gamma_D^*$ fixed as

$$\gamma_D^* = \frac{1}{3} - \nu, \ \nu = 0.4,$$

$$\text{that is, } \gamma_D^* = -0.07, \quad (20)$$

(Housen and Holsapple, 2011). Here, the parameter $\nu$ originates from ~~a~~ coupling parameter $C = r_p U^{\mu} \delta^{\nu}$, a single measure characterizing the impactor in the point source assumption of the impact process, where $\mu$ is another parameter of the coupling parameter. The point source assumption yields the following relationship.

$$\frac{1}{6} < \beta_D = \beta_D^* = \frac{\mu}{2} < \frac{1}{3}, \quad (21)$$

(Holsapple and Schmidt, 1982; Housen and Hosapple, 2003). Eq.19 is rewritten as

$$\pi_D = H_D{}^* \left(\frac{Y}{\delta U^2}\right)^{-\beta_D^*} \pi_4^{\gamma_D^* + \beta_D^*}, \quad (22)$$

from which we obtain

$$\gamma_D^* = \gamma_D - \beta_D^* = \gamma_D - \beta_D. \quad (23)$$

According to Eq. 23 the values of $\gamma_D$ and $\beta_D$ in Eq. 18 collectively give $\gamma_D^* = -0.11$, which is similar to that of $\gamma_D^*$ in Eq. 20 within the uncertainty of $\gamma_D$ and $\beta_D$. The parameters $H_D^*$ and $\beta_D^*$ for highly porous sintered glass bead targets, etc. were also determined using the value of $\gamma_D^*$ fixed as above but using the compressive strength for $Y$ (Okamoto and Nakamura, 2017).

Figure 4b shows a comparison of the data for porous targets according to Eq. 19. We use spall diameter for diameter. Figures 4a and 4b show that the cratering efficiency of porous sedimentary rocks (Suzuki et al., 2012) is intermediate among targets with various porosities studied in the laboratory. Figures 4c and 4d are the compiled results of various targets in terms of scaling parameters, $(10^{-4})^{-\beta_D} H_D$, $(10^{-4})^{-\beta_D^*} H_D^*$, $\beta_D$, and $\beta_D^*$. Fig. 4c illustrates that when the densities of the impactor and target are the same ($\delta = \rho$), the cratering efficiency for the impact of $\frac{Y}{\delta U^2} = \frac{Y}{\rho U^2}$ = $10^{-4}$ generally decreases with porosity from 20 to 30 to low values. In Fig. 4d, the values of $\beta_D$ and $\beta_D^*$ do not show systematic dependence on porosity.

3.4 Crater morphology and volume

As described in Section 3.1, crater shape varies according to porosity. Figure 5 shows the depth-to-diameter ratios of craters on porous targets. We use spall diameter for diameter, except for pumice data. The depth-to-diameter ratio for sandstones is about 0.2, the usual value of lunar craters and craters on non-porous rock targets (Pike et al., 1974; Dohi et al., 2012), whereas craters on more porous targets have larger values of roughly 0.5 for tuff and gypsum targets with porosities of 43% and 50%, respectively. On the other hand, the depth-to-diameter ratios of craters on cement mortar targets are shallower at 0.2, although the porosity is about 40%. Higher porosity does not necessarily result in a larger depth-to-diameter ratio. There is no strong velocity dependence, in contrast to the clear velocity dependence of the depth-to-diameter ratios of microcraters on glass (Mandeville and Vedder, 1971; Hoerz et al., 1975). The velocity dependence of microcraters on glass targets also differs in terms of cratering efficiency as mentioned in Section 3.3. Studies of microcraters on porous targets with impact velocities over 10 km $s^{-1}$ are needed.

Figure 6 shows crater volumes $V_{crater}$ normalized by the cube of the diameter. We use spall diameter for diameter. Previous results for San Marcos gabbro targets (Polanskey and Ahrens, 1990), for which the porosity is 0.2 % (Baud et al., 2014), are also shown. Since crater volume in this case is the sum of the fragmented volume and the volume removed due to compaction,

$$\frac{V_{crater}}{D^3} \geq \frac{V(D)}{D^3} = k_2. \quad (24)$$

Therefore, the vertical axis of Fig. 6 gives the upper limit of $k_2$. The range values in Fig. 6 is similar to or lower than that of 0.075-0.10 for highly porous materials noted previously (Housen and Holsapple, 2011). The crater shape is roughly a trigonal pyramid of diameter $D$ and height $d$ for craters with a depth-to-diameter ratio of less than 0.3, except for the datum for Coconino sandstone.

## 4. SUMMARY

The diameter of crater in the strength regime on a small body is constrained by the mechanical strength of the surface, and the thickness of the regolith layer that covers the surface.

For craters in the strength regime, data from laboratory impact experiments on brittle targets including various porous targets, are now available, and show the following:

1. Spallation can be seen in targets with porosity up to 60%. The ratio of spall diameter to pit diameter is not strongly dependent on porosity and is between 1.5 and 3. The range of the ratio is similar to those found for microcraters on lunar rocks and glass targets in the laboratory.
2. The depth of a crater cavity is a function of the density ratio of the projectile and target. However, it is also dependent on the impact velocity and porosity of the target.
3. The normalized diameters of craters on porous targets tends to decrease with increasing target porosity. An empirical scaling law derived for porous sedimentary rocks (Suzuki et al., 2012) based on conventional scaling laws (Holsapple and Schmidt, 1982) is shown to be a reference for craters on brittle targets, including porous targets of various porosities.

4. The depth-to-diameter ratio of the crater cavity is roughly 0.5 for tuff and gypsum, with porosities of about 43% and 50%, respectively. On the other hand, the ratio is about 0.2 for sandstones and cement mortar, although the porosity of cement mortar is about 40% and similar to that of tuff. No strong velocity dependence is evident in the depth-to-diameter ratio, although the ratio changes with impact velocity for microcraters produced on non-porous glass.
5. Crater shape is roughly a trigonal pyramid for craters with a depth-to-diameter ratio of less than 0.3.

## ACKNOWLEDGMENTS

Two anonymous reviewers were thanked for their careful reading and constructive comments which improved the manuscript much. This research was supported by the Hypervelocity Impact Facility (former facility name: the Space Plasma Laboratory), ISAS, JAXA, Hosokawa Powder Technology Foundation, and a grant-in-aid for scientific research from the Japanese Society for the Promotion of Science (No. 25400453) of the Japanese Ministry of Education, Culture, Sports, Science, and Technology (MEXT).

## REFERENCES

Baldwin, E. C., Milner, D. J., Burchell, D. M. J., Crawford, I. A., 2007. Laboratory impacts into dry and wet sandstone with and without an overlying water layer: Implications for scaling laws and projectile survivability, Meteorit. Planet. Sci., 42, 1905–1914.

Basilevsky, A. T., Lorenz, C. A., Shingareva, T. V., Head, J. W., Ramsley, K. R., Zubarev, A. E., 2014. The surface geology and geomorphology of Phobos, Planet. Space Sci.,

102, 95-118.

Baud, P., Wong, T.-f., Zhu, W., 2014. Effects of porosity and crack density on the compressive strength of rocks. Int. J. Rock Mechanics & Mining Sci., 67, 202-211.

Burchell, M. J., Leliwa-Kopystynski, J., 2010. The large crater on the small Asteroid (2867) Steins. Icarus 210, 707-712.

Buhl, E., Poelchau, M. H., Dresen, G., Kenkmann, T. 2013. Deformation of dry and wet sandstone targets during hypervelocity impact experiments, as revealed from the MEMIN program. Meteoritics & Planet. Sci. 48, 71-86.

Consolmagno, G. J., Britt, D. T., Macke, R. J., 2008. The significance of meteorite density and porosity. Chem. Erde 68, 1–29.

Delbo, M., Libourel, G., Wilkerson, J., Murdoch, N., Michel, P., et al., 2014. Thermal fatigue as the origin of regolith on small asteroids. Nature 508, 233–236.

Dohi, K., Arakawa, M., Okamoto, C., Hasegawa, S., Yasui, M, 2012. The effect of a thin weak layer covering a basalt block on the impact cratering process. Icarus 218, 751-759.

Flynn, G. J., Durda, D. D., Patmore, E. B. Clayton, A. N., Jack, S. J. Lipman, M. D., Strait, M. M., 2015. Hypervelocity cratering and disruption of porous pumice targets: Implications for crater production, catastrophic disruption, and momentum transfer on porous asteroids. Planetary and Space Sci., 107. 64–76.

Fujiwara, A., Kawaguchi, J., Yeomans, D. K., Abe, M., Mukai, T., Okada, T., Saito, J., Yano, H., Yoshikawa, M., Scheeres, D. J., Barnouin-Jha, O., Cheng, A. F., Demura, H., Gaskell, R. W., Hirata, N., Ikeda, H., Kominato, T., Miyamoto, H., Nakamura, A. M., Nakamura, R., Sasaki, S., Uesugi, K., 2006. The Rubble-pile Asteroid Itokawa as Observed by Hayabusa. Science 312, 1330–1334.

Guettler, C., Hirata, N., Nakamura, A. M., 2012. Cratering experiments on the self armoring of coarse-grained granular targets. Icarus 220, 1040–1049.

Harries, D., Yakame, S., Karouji, Y., Uesugi, M., Langenhorst, F., 2016. Secondary submicrometer impact cratering on the surface of asteroid 25143 Itokawa. Earth Planet. Sci. Letters 450, 337-345.

Hiraoka, K., 2008. Experimental Study of the Impact Cratering Process in the Strength Regime. Graduate School of Science and Technology, Kobe University, Doctral dissertation , 1–57.

Hirata, N., Barnouin-Jha, O. S., Honda, C., et al., 2009. A survey of possible impact structures on 25143 Itokawa. Icarus 200, 486–502.

Hoerz, F., Brownlee, D. E., Fechtig, H., Hartung, J. B., Morrison, D. A., Neukum, G., Schneider, E., Vedder, J. F., Gault, D. E., 1975. Lunar microcraters: Implications for the micrometeoroid complex. Planet. Space Sci. 23, 151-172.
Holsapple, K. A., 1993. The scaling of impact processes in planetary sciences. In Annual review of earth and planetary sciences. 21, 333–373.
Holsapple, K. A., Schmidt, R. M., 1982. On the scaling of crater dimensions: 2. Impact processes, J. Geophys. Res., 87, 1849-1870.
Housen, K. R., 1992. Crater ejecta velocities for impacts on rocky bodies. Abstracts of the Lunar and Planetary Science Conference 23, pp. 555-556.
Housen, K. R., Holsapple, K. A., 2003. Impact cratering on porous asteroids. Icarus 163, 102–119.
Housen, K. R., Holsapple, K. A., 2011. Ejecta from impact craters .Icarus 211, 856–875.
Housen, K. R., Holsapple, K. A., 2012. Craters without ejecta. Icarus 219, 297-306.
Housen, K. R., Holsapple, K. A., Voss, M. E., 1999. Compactions as the origin of the unusual craters on the asteroid Mathilde. Nature 402, 155–157.
Housen, K. R., Wilkening, L. L., 1982. Regoliths on small bodies in the solar system. In Annual Rev. Earth and Planet. Sci. 10, 355-376.
Jutzi, M., Michel, P., Hiraoka, K., Nakamura, A. M., Benz, W. 2009. Numerical simulations of impacts involving porous bodies II. Comparison with laboratory experiments, Icarus 201, 802-813.
Kenkmann, T., Wünnemann, K., Deutsch, A., Poelchau, M. H., Schäfer, F., Thoma, K., 2011. Impacts cratering in sandstone: The MEMIN pilot study on the effect of pore water, Meteorit. Planet. Sci., 46, 890–902.
Ko, T.Y., Kemeny, J., 2013. Determination of the subcritical crack growth parameters in rocks using the constant stress-rate test. Int. J. Rock Mech. Min. Sci. 59, 166–178.
Kobayashi, H., Tanaka, H., 2010. Fragmentation model dependence of collision cascades. Icarus 206, 735–746.
Lee, P. Veverka, J., Thomas, P. C., Helfenstein, P., Belton, M. J. S., Chapman, C. R., Greeley, R., Pappalardo, R. T., Sullivan, R., Head, J. W. III, 1996. Ejecta blocks on 243 Ida and on other asteroids. Icarus 120, 87-105.
Lee, S. W., Thomas, P., Veverka, J., 1986. Phobos, Deimos, and the moon - Size and distribution of crater ejecta blocks. Icarus 68, 77–86.
Love, S.G., Hörz, F., Brownlee, D. E., 1993. Target Porosity Effects in Impact Cratering and Collisional Disruption. Icarus 105, 216–224.
Mandeville, J.-C., Vedder, J. F., 1971, Microcraters formed in glass by low density projectiles. Earth and Planetary Science Letters 11, 297-306.
Matsumoto, T., Tsuchiyama, A., Uesugi, K., Nakano, T., Uesugi, M., Matsuno, J., Nagano, T., Shimada, A., Takeuchi, A., Suzuki, Y., Nakamura, T., Nakamura, M., Gucsik, A.,

Nagagki, K., Sakaiya, T., Konda, T., 2016. Nanomorphology of Itokawa regolith particles: Application to space-weathering processes affecting the Itokawa asteroid. Geochimica et Cosmochimica Acta 187, 195-217.

Melosh, H. J., 1989. Impact Cratering: A Geologic Process. Oxford University Press, New York, (Oxford Monographs on Geology and Geophysics, No. 11).

Michikami, T., Hagermann, A., Morota, T., Haruyama, J., Hasegawa, S., 2017. Oblique impact cratering experiments in brittle targets: Implications for elliptical craters on the Moon. Planet. Space Sci. 135, 27–36.

Michikami, T., Moriguchi, K., Hasegawa, S., Fujiwara, A., 2007. Ejecta velocity distribution of impact cratering experiments on porous and low strength targets. Planet. Space Sci. 55, 70–88.

Michikami, T., Nakamura, A. M., Hirata, N., Gaskell, R. W., Nakamura, R., Honda, T., Honda, C., Hiraoka, K., Saito, J., Demura, H., Ishiguro, M., Miyamoto, H. 2008. Size-frequency statistics of boulders on global surface of asteroid 25143 Itokawa. Earth Planets Space 60, 13–20.

Morrison, D. A., MaKay, D. S., Fruland, R. M., Moore, H. J., 1973. Microcraters on Apollo 15 and 16 rocks. Proc. Fourth Lunar Sci. Conf., Supplement 4, Geochimica et Cosmochimica Acta 3, 3235-3253.

Nakamura, A., Hiraoka, K., Yamashita, Y., Machii, N., 2009. Collisional disruption experiments of porous targets. Planet. Space Sci. 57, 111–118.

Nakamura, E., Makishima, A., Moriguti, T., Kobayashi, K., Tanaka, R., Kunihiro, T., Tsujimori, T., Sakaguchi, C., Kitagawa, H., Ota, T., Yachi, Y., Yada, T., Abe, M., Fujimura, A., Ueno, M., Mukai, T., Yoshikawa, M., Kawaguchi. J., 2012. Space environment of an asteroid preserved on micrograins returned by the Hayabusa spacecraft. Proc. Natl. Acad. Sci. U.S.A. 109, E624–E629.

Okamoto, T., Nakamura, A. M., 2017. Scaling of Impact-generated Cavity-size for Highly Porous Targets and Its Application to Cometary Surfaces. Icarus 292, 234-244.

Okamoto, T., Nakamura, A.M., Hasegawa, S., 2015. Impact experiments on highly porous targets: Cavity morphology and disruption thresholds in the strength regime. Planetary and Space Science 107, 36–44.

Okamoto, T., Nakamura, A. M., Hasegawa, S., Kurosawa, K., Ikezaki, K., Tsuchiyama, A., 2013. Impact experiments of exotic dust grain capture by highly porous primitive bodies. Icarus 224, 209-217.

Pike, R.J., 1974. Depth/diameter relations of fresh lunar craters: Revision from spacecraft data. Geophysical Research Letters 1, 291–294.

Poelchau, M. H., Kenkmann, T., Thoma, K., Hoerth, T., Dufresne, A., Schäfer, F., 2013.

The MEMIN research unit: Scaling impact cratering experiments in porous sandstones. Meteorit. Planet. Sci., 48, 8-22.
Poelchau, M. H., Kenkmann, T., Hoerth, T., Schäfer, F., Rudolf, M., Thoma, K., 2014. Impact cratering experiments into quartzite, sandstone and tuff: The effects of projectile size and target properties on spallation. Icarus 242, 211–224.
Polanskey, C. A., Ahrens, T. J., 1990. Impact spallation experiments: Fracture patterns and spall velocities. Icarus 87, 140-155.
Suzuki, A. Hakura, S. Hamura, T. Hattori, M. Hayama, R. Ikeda, T. Kusuno, H. Kuwahara, H. Muto, Y. Nagaki, K. Niimi, R. Ogata, Y. Okamoto, T. Sasamori, T. Sekigawa, C. Yoshihara, T. Hasegawa, S. Kurosawa, K., Kadono, T., Nakamura, A. M., Sugita, S., Arakawa, M., 2012. Laboratory experiments on crater scaling-law for sedimentary rocks in the strength regime. J. Geophys. Res. 117, E08012, doi:10.1029/2012JE004064.
Thomas, P. C., Veverka, J., Robinson, M. S., Murchie, S., 2001. Shoemaker crater as the source of most ejecta blocks on the asteroid 433 Eros. Nature 413, 394–396.
Tomeoka, K., Kiriyama, K., Nakamura, K., Yamahana, Y., Sekine, T., 2003. Interplanetary dust from the explosive dispersal of hydrated asteroids by impacts. Nature 423, 60-62.
Veverka, J., Thomas, P., Harchi, A., Clark, B., Bell, J. F., Carcich, B., Joseph, J., Murchie, S., Izenberg, N., Chapman, C., Merline, W., Malin, M., McFadden, L., Robinson, M., 1999. NEAR Encounter with Asteroid 253 Mathilde: Overview. Icarus 140, 3–16.
Winkler, R., Poelchau, M. H., Moser, S., Kenkmann, T., 2016. Subsurface deformation in hypervelocity cratering experiments into high-porosity tuffs. Meteoritics & Planet. Sci. 51, 1849-1870.
Yamamoto, S., Mukai, T., 1998. Dust production by impacts of interstellar dust on Edgeworth-Kuiper Belt objects. Astron. & Astrophys. 329, 785-791.
Yasui, M., Arakawa, M., Hasegawa, S., Fujita, Y., Kadono, T., 2012. In situ flash X-ray observation of projectile penetration processes and crater cavity growth in porous gypsum target analogous to low-density asteroids. Icarus 221, 646–657.

Table 1 Porous targets used in high-velocity impact cratering experiments in the strength regime.

| Target material | | | | Projectile material, diameter (mm) | Impact speed (km $s^{-1}$) | Reference |
|---|---|---|---|---|---|---|
| Material | Density[a] (kg $m^{-3}$) | Porosity[a] (%) | Strength[a, b] (MPa) | | | |
| Weak cemented basalt [c] | 2600 | 23 [c] | 0.68 (C)<br>*0.090* [d] (T) | Al, 6.35 | 1.86 | 1 |
| | 2600 | 23 [c] | 10 (C)<br>*0.45* [d] (T) | | | |
| Sand–-perlite–fly ash mixture | 690-1750 | 35-72 [e] | ~0.02 (C) [f]<br>~0.004 (T) [f] | High-density polyethylene, 12.2∅ × 12.2 (cylinder) | 1.80-1.94 | 2 |
| Sintered glass bead | 1000 | 60 | 0.2 (C) | Soda lime glass, 1.588 | 5.965-6.056 | 3 |
| | 1500 | 39 | 0.79 (C) | | | |
| | 1580 | 37 | 4.4 (C) | | | |
| | 2440 | 5 | 28 (C) | | | |
| | 410 | 84 | - | Alumina, ~1 | 1.22-4.47 | 4 |
| | 1019, 1046 | 59, 58 | 0.8 (C) | | | |
| | 1421-1424 | 43~~43~~ | 0.53 (C) | | | |
| | 1479-1547 | 41-38 | 5.5 (C) | | | |
| | 1727, 1796 | 31, 28 | 34.2 (C) | | | |
| | 2394-2269 | 4-9 | 245 (C) | | | |
| | 1490 | 40 | 1.9 (C),<br>0.37 (T) | Nylon, 7;<br>Glass, 0.94 and 3.2;<br>Alumina, 1.18;<br>SUS-440C, 1.6 | 1.90-3.46 | 5 |
| | 1590 | 37 | 12 (C),<br>1.8 (T) | | | |
| | ~1700 | ~32 | 41 (C) | | | |
| | 140 | 94 | 0.47 (C) | Polystyrene, 1.1;<br>Nylon, 3.2; | 2.26-7.17 | 6 |
| | 165 | 93 | 0.025 (C) | | | |
| | 340 | 87 | 1.43 (C) | Al, 1.0;<br>Basalt , 3$\emptyset \times$ 2 (cylinder);<br>Ti ,1.0 | | |
| Mixture of sintered glass bead and silicate grain | 1690 | 34 | 15 (C),<br>2.0 (T) | Glass, 0.94 and 3.2;<br>SUS-440C, 1.6 | 2.51-3.53 | 5 |
| | 1800 | 32 | 18 (C),<br>3.5 (T) | | | |
| Cement mortar | 1550 | ~40 | 3.2 (C),<br>0.83 (T) | Nylon, 7.14 | 2.44 | 7 |
| Gypsum | 1100 | ~50 | 15.6 (C),<br>2.52 (T) | Nylon, 3.2;<br>Al, 3.2;<br>Stainless steel, 1.6 | 2.12-6.35 | 6, 8 |
| Coconino sandstone | 1780 | ~23 | ~74 (C),<br>*118.0 (C)* [g]<br>*6.4 (T)* [g],<br>*12.7 (T)* [h] | Stainless steel 420, 1 | 5.10 | 9 |
| Sandstone | 2180 | ~17 | 90 (C)<br>*15.4 (T)* [h] | Stainless steel 420, 1 | 5.03 | 9 |
| Seeberger sandstone | 2050 | 23 | 67.3 (C)<br>4.1 (T) | D290-1 steel, 2.5, 10 and 12;<br>AISI 4130 steel, 10;<br>Iron meteorite, 10 | 2.50-5.34 | 10, 11 |
| Pakistan sedimentary rock (PS) | 2240 | ~17 | 4.6 (T) | Nylon, 1$\phi \times$ 1 (cylinder) and 3.2;<br>Soda lime glass, 1.1;<br>Alumina, 1.0;<br>Ti, 1;<br>Stainless steel, 1.0;<br>Cupper, 1.0;<br>WC, 1.1 | 2.0-6.9 [i] | 12 |
| China sedimentary rock (CS) | 2240 | ~15 | 3.0 (T) | Nylon, 7.1 | 2.2, .4.0 | 12 |
| Weibern tuff | 1420 | 43 | 12.3 (C)<br>1.64 (T) | D290-1 steel, 2.5 and 12 | 4.76-5.57 | 11 |
| Pumice (Ito ignimbrite from Aira caldera) | 590 | 74 | 5.1 (C)<br>1.0 (T) [j] | Nylon, 3.2 | 3.58-7.19 | 6 |

[a] Each density corresponds to each porosity and strength.

[b] C and T denote compressive and tensile strength, respectively.

[c] Both targets are referred to as “weak cemented basalt” in Housen and Holsapple (2011). The higher strength comes from adding slightly more binding agent, which did not have a significant effect on the porosity (K. Housen, private communication in 2017).

[d] Tensile strength was estimated using an empirical relationship (see, ref. 1 and references therein).

[e] Data for pure sand and pure perlite are not included.

[f] The least porous targets were twice as strong.

[g] Ko and Kemeny (2013).

[h] Estimated and used in plots in Suzuki et al. (2012).

[i] Data for lower velocity shots with impact velocities from 0.82 to 1.04 km $s^{-1}$ are not referred to in this review.

[j] Jutzi et al. (2009).

1 Housen (1992), 2 Housen and Holsapple (2003), 3 Love et al. (1993), 4 Michikami et al. (2007), 5 Hiraoka (2008), 6 Okamoto and Nakamura (2017), 7 Michikami et al. (2017), 8 Yasui et al. (2012), 9 Baldwin, et al. (2007), 10 Poelchau et al. (2013), 11 Poelchau et al. (2014), 12 Suzuki et al. (2012)

Figure captions

Fig. 1 Diameter range of craters in the strength regime which is restricted below the gravity-strength transition indicated by the thick solid line ($Y$=0.1 MPa, $\rho$=1500 kg $m^{-3}$, $k_3$=1 in Eq. 3) and the thick dashed line ($Y$=0.45 MPa, $\rho$=2600 kg $m^{-3}$, $k_3$=1 in Eq. 3) and the maximum crater diameter, $D_{max}$, and above the thickness of regolith. The thin solid line and thin dashed line indicate the average regolith thickness owing to the largest crater for a material of $Y$=0.1 MPa, $\rho$=1500 kg $m^{-3}$, and the empirical function of escape fraction of ejecta presented in Michikami et al. (2008) and for the one of $Y$=0.45 MPa, $\rho$=2600 kg $m^{-3}$, and Eq. 10 with the parameter set of weak cemented basalt, $C_s$=0.122 and $\beta_s$=1.38, respectively, although the value of $C_s$ was derived based on an advanced ejecta model (Housen and Holsapple, 2011) and not on a simple power-law assumed here. The thin dotted line shows the upper limit derived by assuming complete re-accumulation of the ejecta.

Fig. 2 Spall diameter to pit diameter versus target porosity. Microcraters on glass target (Mandeville and Vedder, 1971), millimeter-centimeter craters on sintered glass bead (Michikami et al., 2007; Hiraoka, 2008), mixture of sintered glass bead and silicate

Weibern tuff (Poelchau et al., 2014), Pakistan sedimentary rock (PS) and China sedimentary rock (CS) (Suzuki et al., 2012). Filled marks are those of impact velocity less than 4 km $s^{-1}$, whereas open marks and crosses are those of higher impact velocities. The line corresponds to an empirical relationship shown by Eq. 14 obtained for the highly porous targets of foamed polystyrene, sintered glass bead, gypsum, and pumice (Okamoto and Nakamura, 2017). The data for gypsum and pumice used in the fitting of the empirical relationship are also shown in light blue ($U$ > 4 km $s^{-1}$) and dark blue ($U$ < 4 km $s^{-1}$).

Fig. 4 Normalized crater diameters of porous targets. (a) The solid line shows an empirical relationship (Eq. 16 along with 18) obtained for porous sedimentary rocks (PS and CS) (Suzuki et al., 2012). We adopted a tensile strength of 6.4 MPa for Coconino sandstone. Data for microcraters on glass (Mandeville and Vedder, 1971) and centimeter-size craters on basalt are also shown (Dohi et al., 2012). For glass, a tensile strength of 200 MPa is assumed. (b) The value of $\nu$ was assumed to be 0.4. Lines show empirical relationship for weak cemented basalt, fly ash-sand mixture, perlite-sand mixture targets and rock (Housen and Holsapple, 2011). (c) Scaling parameters $(10^{-4})^{-\beta_D} H_D$ (open marks) and $(10^{-4})^{-\beta_D^*} H_D^*$ (for rock and filled marks for others) and (d) $\beta_D$ (open marks) and $\beta_D^*$ (for rock and filled marks for others) are shown versus porosity. The value of $\beta_D^*$ for highly porous targets (sintered glass bead targets, sgb, and foamed polystyrene targets, fp) and an empirical relationship between $\beta_D^*$ and porosity (Okamoto and Nakamura, 2017) are also shown.

Fig. 5 Depth-to-diameter ratio of craters on porous targets versus (a) impact velocity, and (b) the projectile to target density ratio $\delta/\rho$. The data for microcraters on glass target are also shown.

Fig. 6 Crater cavity volume normalized by the cube of diameter versus (a) the target porosity and (b) depth-to-diameter ratio (*d/D*) of the crater. The line in (b) is a reference line of a trigonal pyramid of diameter *D* and height *d*. Data for lower velocity shots of San Marcos gabbro with impact velocities of 0.89 and 1.01 km $s^{-1}$ are not included in the figures.

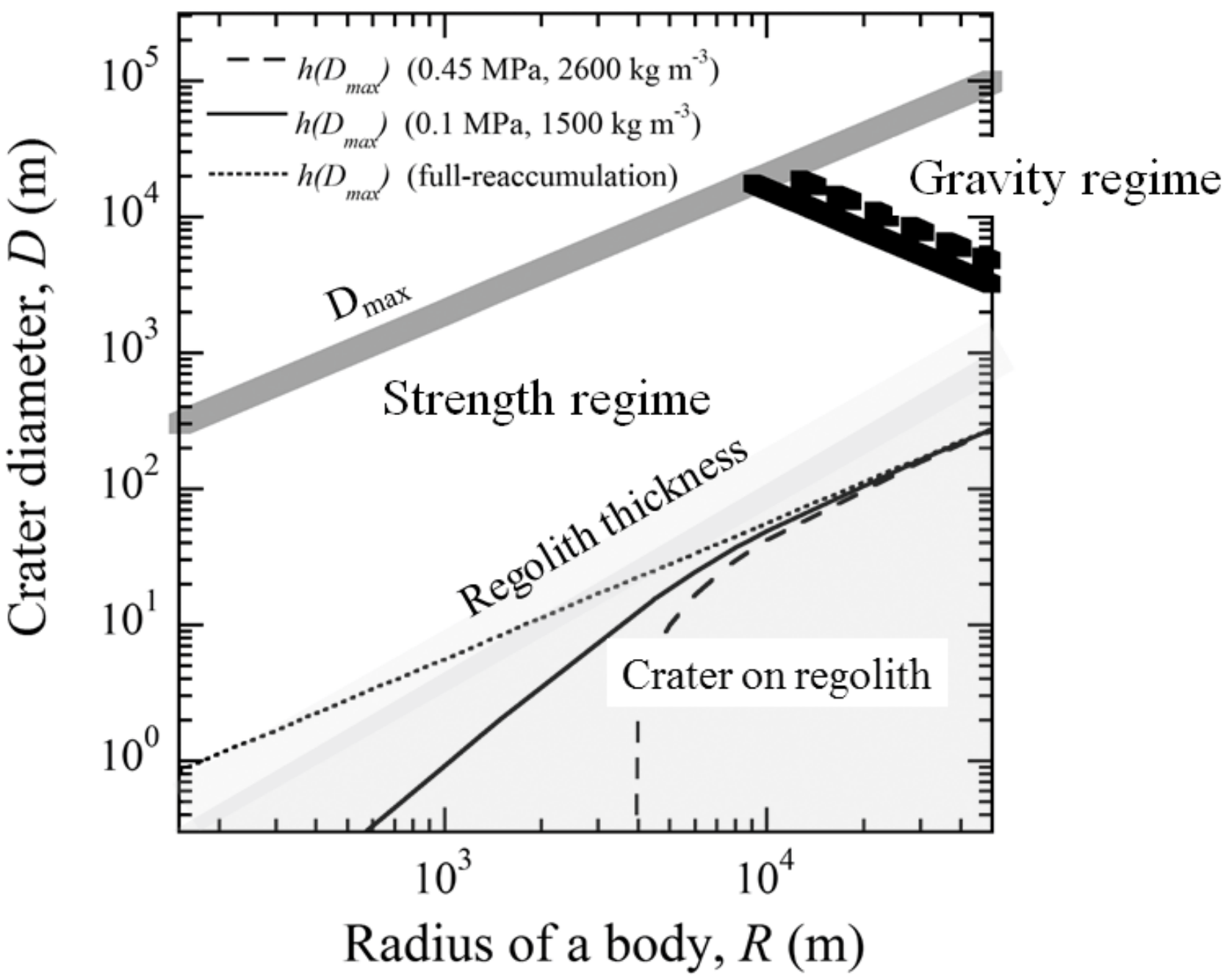

Fig. 1 Diameter range of craters in the strength regime which is restricted below the gravity-strength transition indicated by the thick solid line ($Y$=0.1 MPa, $\rho$=1500 kg m$^{-3}$, $k_3$=1 in Eq. 3) and the thick dashed line ($Y$=0.45 MPa, $\rho$=2600 kg m$^{-3}$, $k_3$=1 in Eq. 3) and the maximum crater diameter, $D_{max}$, and above the thickness of regolith. The thin solid line and thin dashed line indicate the average regolith thickness owing to the largest crater for a material of $Y$=0.1 MPa, $\rho$=1500 kg m$^{-3}$, and the empirical function of escape fraction of ejecta presented in Michikami et al. (2008) and for the one of $Y$=0.45 MPa, $\rho$=2600 kg m$^{-3}$, and Eq. 10 with the parameter set of weak cemented basalt, $C_s$=0.122 and $\beta_s$=1.38, respectively, although the value of $C_s$ was derived based on an advanced ejecta model (Housen and Holsapple, 2011) and not on a simple power-law assumed here. The thin dotted line shows the upper limit derived by assuming complete re-accumulation of the ejecta.

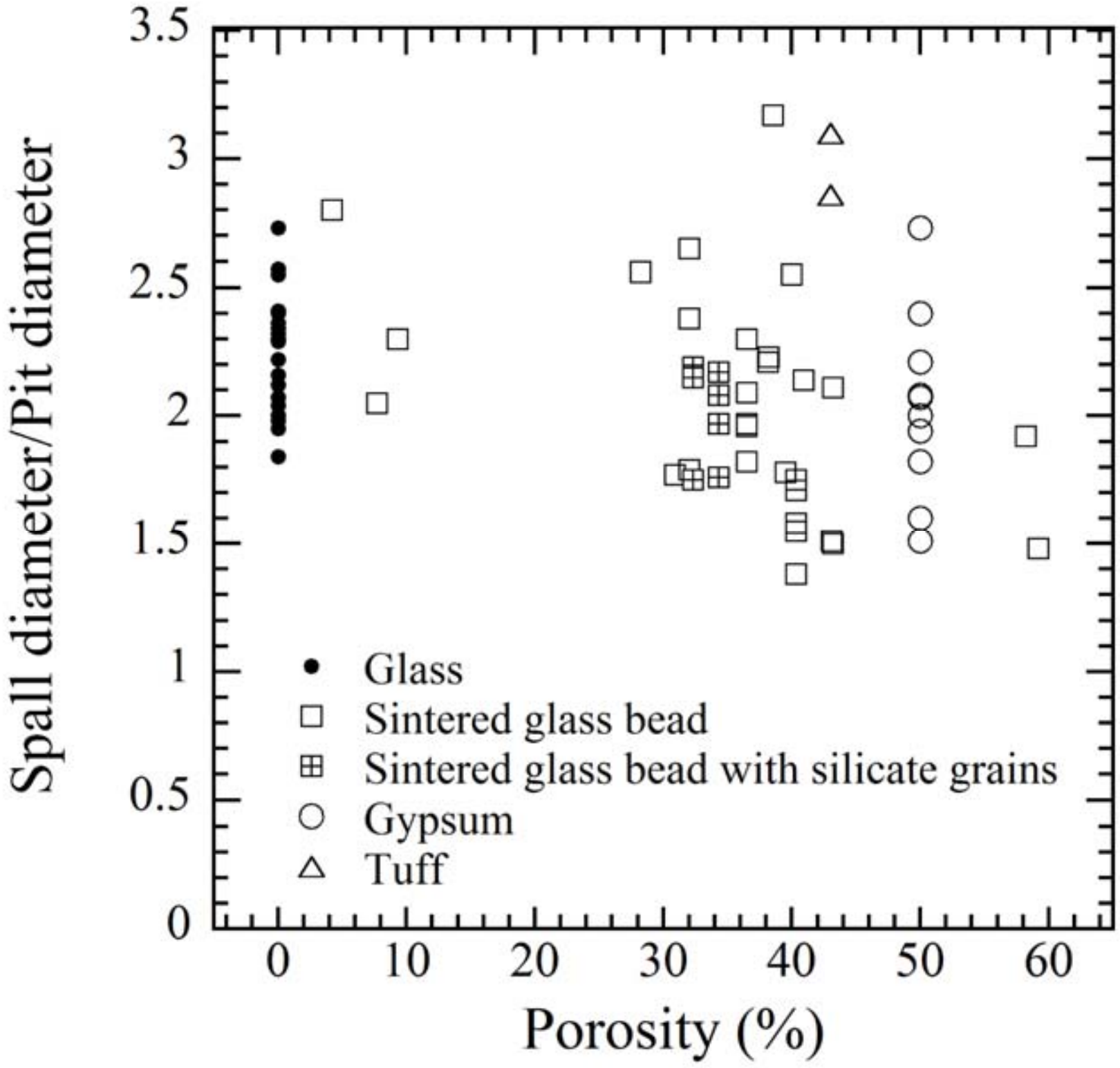

Fig. 2 Spall diameter to pit diameter versus target porosity. Microcraters on glass target (Mandeville and Vedder, 1971), millimeter-centimeter craters on sintered glass bead (Michikami et al., 2007; Hiraoka, 2008), mixture of sintered glass bead and silicate grain (Hiraoka, 2008), gypsum (Yasui et al., 2012), and Weibern tuff (Winkler et al., 2016) targets.

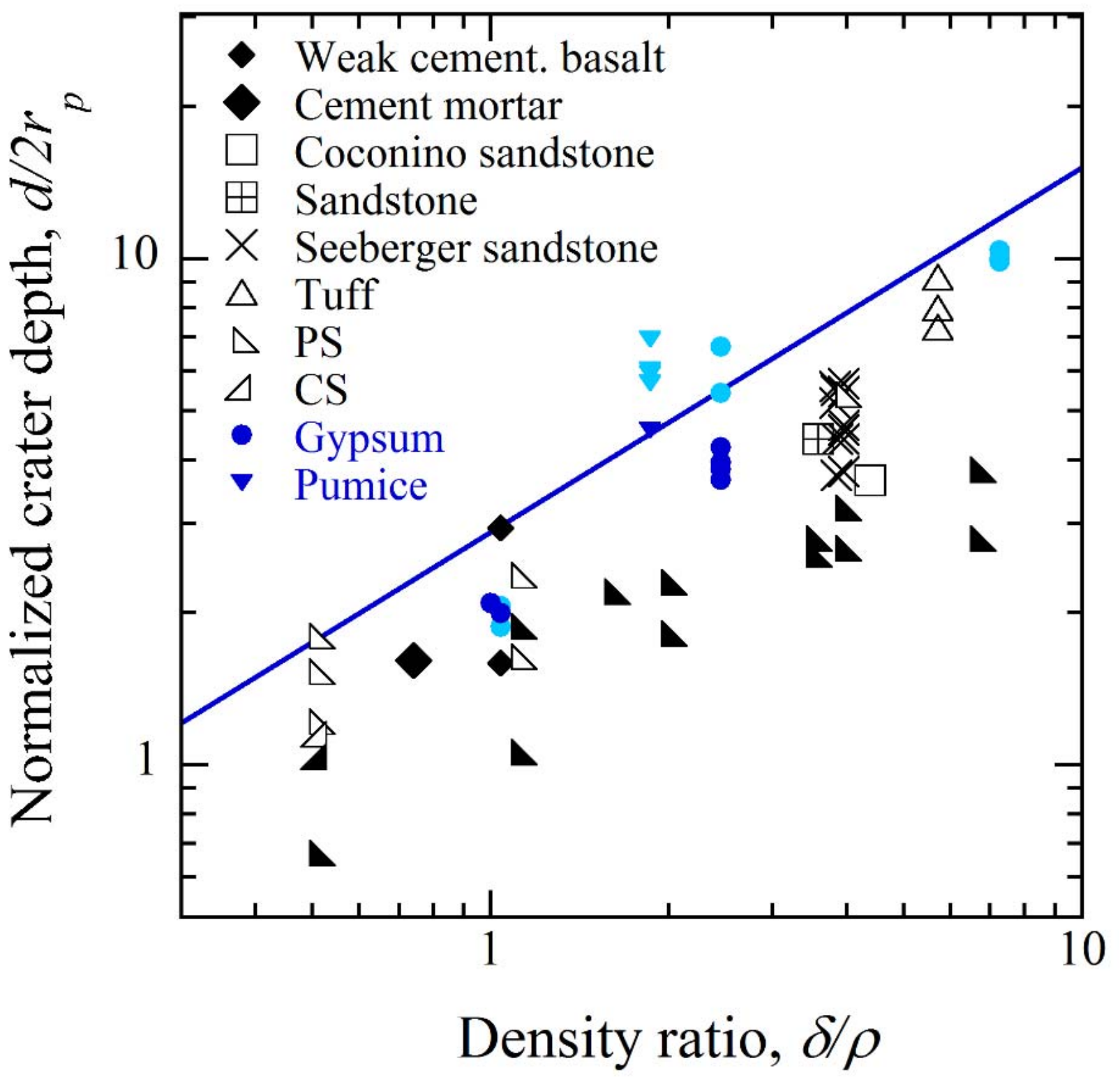

Fig. 3 Normalized crater depth versus the ratio of projectile density to target density for weak cemented basalt (Housen, 1992), cement mortar (Michikami et al., 2017), Coconino sandstone and sandstone (Baldwin et al., 2007), Seeberger sandstone and Weibern tuff (Poelchau et al., 2014), Pakistan sedimentary rock (PS) and China sedimentary rock (CS) (Suzuki et al., 2012). Filled marks are those of impact velocity less than 4 km $s^{-1}$, whereas open marks and crosses are those of higher impact velocities. For a cylindrical projectile of one of PS shots, the height of the cylinder is used instead of $2r_p$ for the normalization of crater depth. The line corresponds to an empirical relationship shown by Eq. 14 obtained for the highly porous targets of foamed polystyrene, sintered glass bead, gypsum, and pumice (Okamoto and Nakamura, 2017). The data for gypsum and pumice used in the fitting of the empirical relationship are also shown in light blue ($U > 4$ km $s^{-1}$) and dark blue ($U < 4$ km $s^{-1}$).

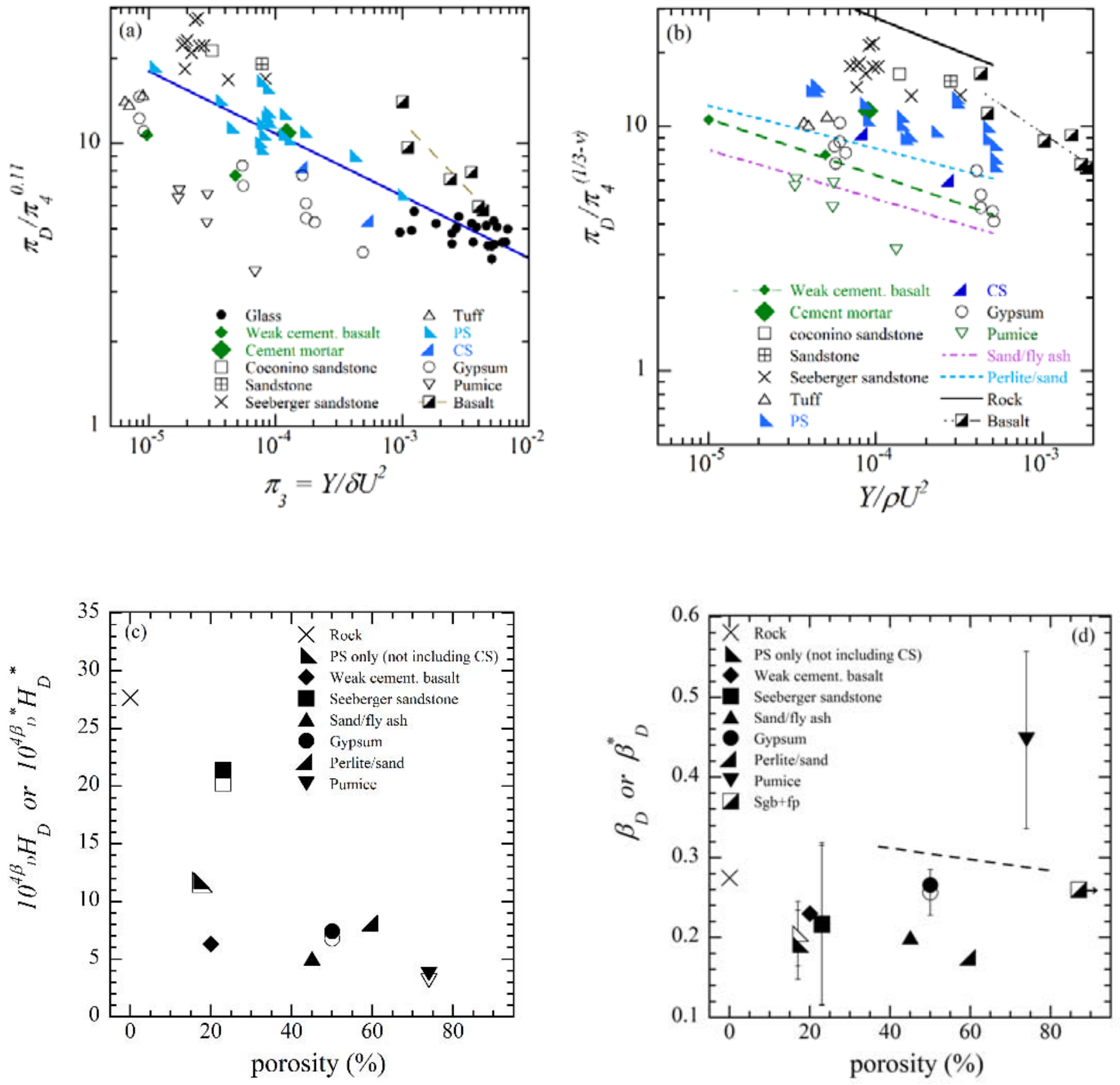

Fig. 4 Normalized crater diameters of porous targets. (a) The solid line shows an empirical relationship (Eq. 16 along with 18) obtained for porous sedimentary rocks (PS and CS) (Suzuki et al., 2012). We adopted a tensile strength of 6.4 MPa for Coconino sandstone. Data for microcraters on glass (Mandeville and Vedder, 1971) and centimeter-size craters on basalt are also shown (Dohi et al., 2012). For glass, a tensile strength of 200 MPa is assumed. (b) The value of $\nu$ was assumed to be 0.4. Lines show empirical relationship for weak cemented basalt, fly ash-sand mixture, perlite-sand mixture targets and rock (Housen and Holsapple, 2011). (c) Scaling parameters $(10^{-4})^{-\beta_D} H_D$ (open marks) and $(10^{-4})^{-\beta_D^*} H_D^*$ (for rock and filled marks for others) and (d) $\beta_D$ (open marks) and $\beta_D^*$ (for rock and filled marks for others) are shown versus porosity. The value of $\beta_D^*$ for highly porous targets (sintered glass bead targets, sgb, and foamed polystyrene targets, fp) and an empirical relationship between $\beta_D^*$ and porosity (Okamoto and Nakamura, 2017) are also shown.

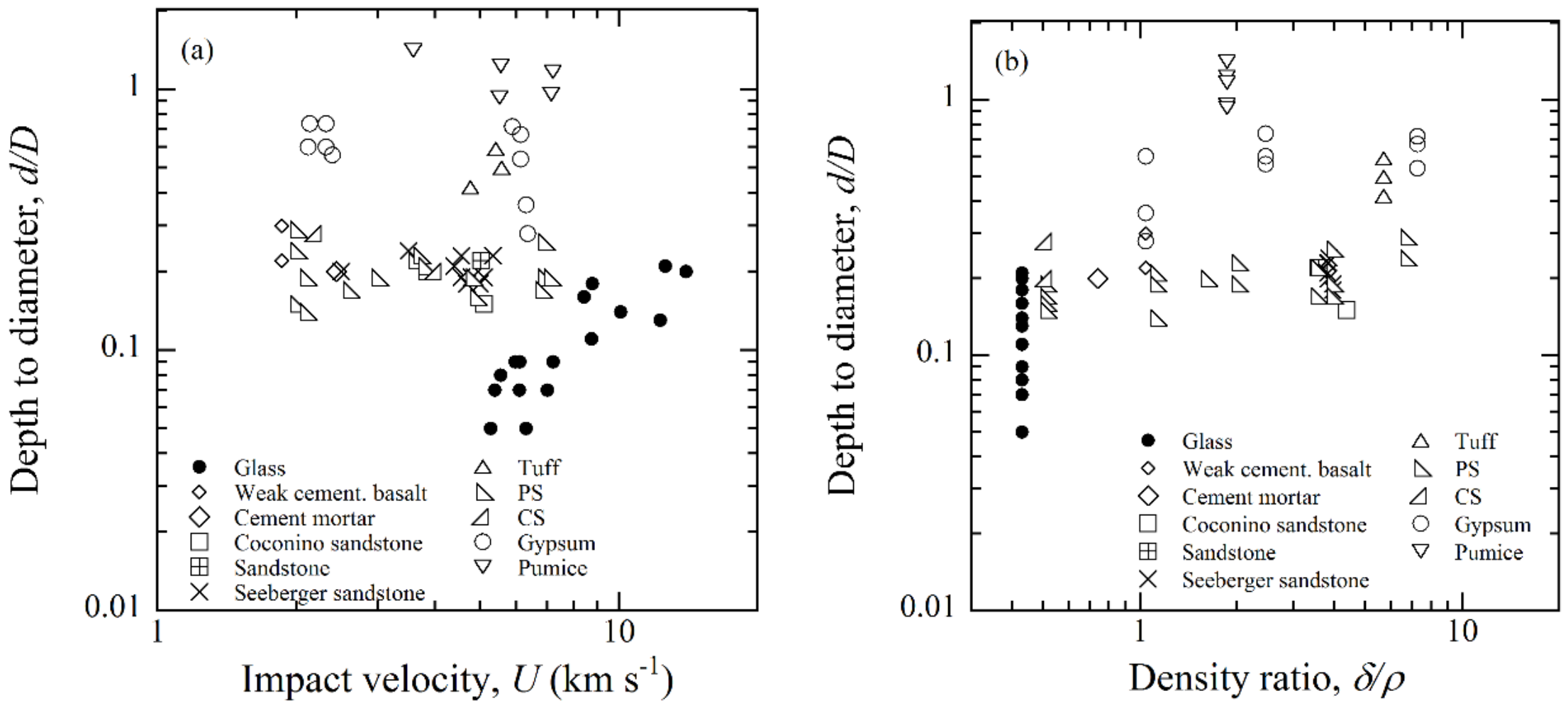

Fig. 5 Depth-to-diameter ratio of craters on porous targets versus (a) impact velocity, and (b) the projectile to target density ratio $\delta/\rho$. The data for microcraters on glass target are also shown.

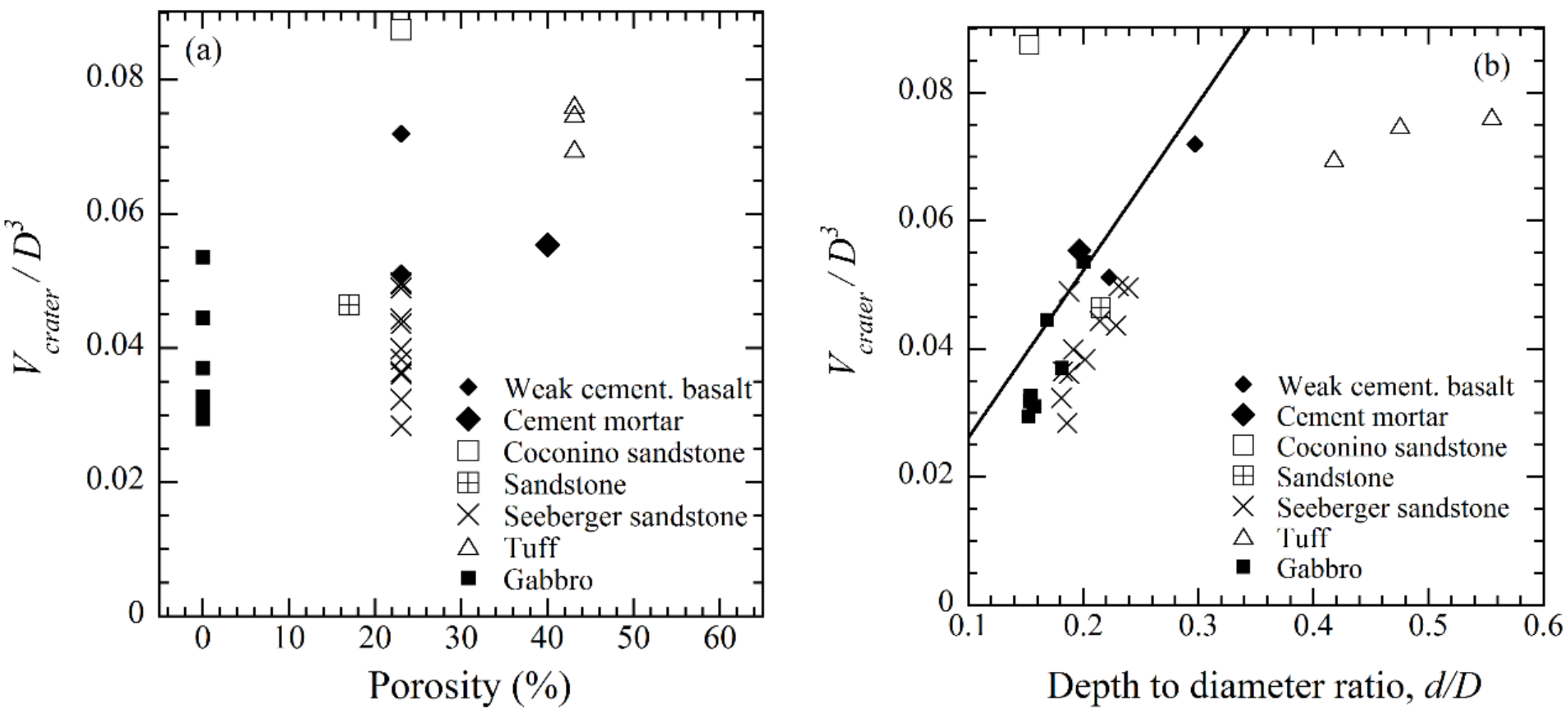

Fig. 6 Crater cavity volume normalized by the cube of diameter versus (a) the target porosity and (b) depth-to-diameter ratio (*d/D*) of the crater. The line in (b) is a reference line of a trigonal pyramid of diameter *D* and height *d*. Data for lower velocity shots of San Marcos gabbro with impact velocities of 0.89 and 1.01 km $s^{-1}$ are not included in the figures.